\documentclass[preprintnumbers,aps,11pt,floatfix,nofootinbib,showpacs,superscriptaddress]{revtex4}

\usepackage{subfigure}
\usepackage{amssymb}
\usepackage{graphicx}
\usepackage{amsmath}

\usepackage[british]{babel}
\usepackage{graphicx}

\long\def\del #1 \enddel { }

\usepackage{epsfig}

\newcommand{\gammatwo}{\Gamma^{(2)}}

\newcommand{\ZA}{Z_{A,k}}

\newcommand{\Ofield}{\mathop{\bigg\vert_{A=\bar{A}=0}}_{g=\bar{g}=\eta}}

\newcommand{\step}{\vspace{.5em}}

\def\di{\displaystyle}

\def\bg{\begin{eqnarray}\begin{array}{rcl}\displaystyle}
\def\eg{\end{array} &\di    &\di   \end{eqnarray}}
\def\bm#1{\begin{eqnarray}\begin{array}{#1}\di}
\def\bmo#1{\begin{eqnarray*}\begin{array}{#1}\di}
\def\bml#1#2{\begin{eqnarray}\begin{array}{#1}\label{#2}\di}
\def\bgo{\begin{eqnarray*}\begin{array}{rcl}\displaystyle}
\def\ego{\end{array} &\di    &\di \nonumber  \end{eqnarray*}}

\def\btensor#1#2{\renew\left#1\begin{array}{#2}\di}
\def\brtensor#1#2#3{\ren#3\left#1\begin{array}{#2}}
\def\botensor#1#2{\renew\left#1\begin{array}{#2}}
\def\etensor#1{\end{array}\right#1}

\def\eq#1{(\ref{#1})}

\def\tr{{\rm tr}}

\def\s0#1#2{\mbox{\small{$ \frac{#1}{#2} $}}}
\def\0#1#2{\frac{#1}{#2}}

\def\ren#1{\renewcommand{\arraystretch}{#1}}

\def\renew{\renewcommand{\arraystretch}{1}}

\def\gYM{ g_{S}}
\def\gYMk{ g_{S,k}}

\def\gN{ g_{N}}
\def\gNk{ g_{N}}

\def\bea{\arraycolsep .1em \begin{eqnarray}}
\def\eea{\end{eqnarray}}
\def\Tr{{\rm Tr}}

\def\eq#1{(\ref{#1})}

\def\s0#1#2{\mbox{\small{$ \frac{#1}{#2} $}}}
\def\0#1#2{\frac{#1}{#2}}

\def\grgl{\:\hbox to -0.2pt{\lower2.5pt\hbox{$\sim$}\hss}{\raise3pt\hbox{$>$}}\:}
\def\klgl{\:\hbox to -0.2pt{\lower2.5pt\hbox{$\sim$}\hss}{\raise3pt\hbox{$<$}}\:}

\begin{document}

\title{Asymptotic freedom of Yang-Mills theory with gravity}
 \vspace{1.5 true cm}
 
\author{Sarah Folkerts\footnote{Present address: Arnold Sommerfeld Center, Ludwigs-Maximilians-Universit\"at, Theresienstr. 37, 80333 M\"unchen, Germany}}
\affiliation{Institut f. Theoretische Physik, Universit\"at Heidelberg, Philosophenweg 16, 69120 Heidelberg, Germany}
\author{Daniel F. Litim}
\affiliation {Department\ of Physics \& Astronomy, University of Sussex,
  Brighton, BN1 9QH, U.K.}
\author{Jan M. Pawlowski}
\affiliation{Institut f. Theoretische Physik, Universit\"at Heidelberg, Philosophenweg 16, 69120 Heidelberg, Germany}
\affiliation{ExtreMe Matter Inst. EMMI, GSI, Planckstr.~1, 64291 Darmstadt, Germany}

%04.60.-m Quantum gravity 
%11.10.Hi Renormalization group evolution of parameters 
%11.15.Tk Other nonperturbative techniques

\begin{abstract} {We study the 
%high energy 
behaviour of Yang-Mills
    theory under the inclusion of gravity.  In the weak-gravity limit,
    the running gauge coupling receives no contribution from the
    gravitational sector, if all symmetries are preserved. This holds
    true with and without cosmological constant.  We also show that
    asymptotic freedom persists in general field-theory-based gravity
    scenarios including gravitational shielding as well as
    asymptotically safe gravity.  }
\end{abstract}

\maketitle

\pagestyle{plain}
\setcounter{page}{1}
\noindent 
\section{ Introduction}

Asymptotic freedom of Yang-Mills theories
\cite{Gross:1973id,Politzer:1973fx} -- the weakening of the strong
force at asymptotically short distances -- is a central
characteristics of the Standard Model. Within a renormalisation group
language, asymptotic freedom is signaled by a trivial ultraviolet
fixed point with vanishing Yang-Mills coupling.  This pattern is at the root of 
the existence of Yang-Mills theory as an asymptotic  perturbative  series
valid for high energies, i.e.~energies much larger than the dynamical strong-coupling scale, $\Lambda_{\rm QCD}$,
and makes the high energy behaviour accessible with weak coupling
methods. It is well-known, however, that this feature can be
destabilized once the couplings to matter degrees of freedom are taken
into account, leading to constraints for the latter to ensure
asymptotic freedom within the SM and its various extensions. \step

Less is known about the stability of asymptotic freedom under the
inclusion of gravity.  Recently, quantum gravitational corrections to
the Yang-Mills coupling have been studied perturbatively to one-loop
order in Newton's coupling
\cite{Robinson:2005fj,Pietrykowski:2006xy,Toms:2007sk,%
  Ebert:2007gf,Tang:2008ah,Toms:2010vy}, by
treating gravity as an effective theory amended by an ultraviolet
cutoff of the order of the Planck scale \cite{Donoghue:1993eb} (see
\cite{Deser:1974nb,Deser:1974cz,Deser:1974xq} for earlier work). The
strength of this technique is that quantum gravitational fluctuations
can be computed in the low energy regime without the knowledge of the
underlying quantum theory for gravity.  Asymptotic freedom remains
intact in these studies and extensions thereof including compact extra
dimensions \cite{Ebert:2008ux,Wu:2008ph} and a cosmological constant
\cite{Toms:2008dq}. 
These one-loop results have recently been confirmed beyond 
perturbation theory in \cite{Daum:2009dn}.\step
 
The findings open up a number of farther reaching questions.  The
one-loop graviton contribution to the Yang-Mills $\beta$-function
comes out regularisation- and gauge-fixing dependent, which raises the
question about its universality.  It has also been argued that the
one-loop coefficient is not invariant under re-parameterisations in
field space, and hence not a measurable quantity \cite{Ellis:2010rw}.
For specific gauges or regularisations a vanishing one-loop
coefficient has been found, and a computation beyond the one-loop
level becomes mandatory. Also, the applicability of the effective
theory is confined to the regime of weak gravity and strong coupling
effects will become important and should be taken into account for
energies approaching the Planck scale.  Finally, at Planck-scale
energies and beyond, an effective theory description is superseded by
a fundamental theory for gravity, and the ultimate fate of asymptotic
freedom for Yang-Mills theory then depends on the ultraviolet
completion for gravity.\step
 
In this Letter, we study Yang-Mills
theory coupled to gravity with the help of the functional
renormalisation group
\cite{Wetterich:1992yh,Litim:1998nf,Gies:2006wv,Pawlowski:2005xe} (see
\cite{Daum:2009dn} for a related study). This technique is based on
the infinitesimal integrating-out of momentum modes from a path
integral representation of quantum field theory by means of an
infrared momentum cutoff. The renormalisation group flow links the
fundamental theory with the corresponding quantum effective action at
low energies.  A particular strength of this method is its
flexibility, allowing for perturbative and non-perturbative
approximations.  In the past, it has been applied to Yang-Mills theory
\cite{Litim:1998nf,Litim:2002ce,Gies:2002af,Gies:2006wv,%
  Pawlowski:2003hq,Pawlowski:2005xe,Reuter:1997gx,Litim:1998qi,%
  Freire:2000bq,Pawlowski:2001df,Pawlowski:2003sk,Litim:2002xm,Litim:2002hj}
and gravity
\cite{Niedermaier:2006ns,Percacci:2007sz,Litim:2008tt,Reuter:1996cp,%
  Souma:1999at,Lauscher:2001ya,Percacci:2002ie,Litim:2003vp,%
  Fischer:2006fz,Codello:2007bd,Benedetti:2009gn} individually, both
at weak and strong coupling, which makes it an ideal tool for the
study of the coupled system \cite{Folkerts:Dipl2009}. \step

In the weak gravity limit, we analyse the gravitational corrections to
Yang Mills theory for general regularisations and backgrounds.  Gauge
and diffeomorphism invariance of the effective action is achieved
using the background field method. We evaluate the background field
independence of our results by extending earlier explicit results
within gauge Yang Mills theory \cite{Litim:2002ce} to quantum gravity.
In this general setting the gravitational contribution to the running
gauge coupling is computed at one-loop. The result encompasses all
previous studies, also including a cosmological constant.  We also
address asymptotic freedom in the limit where the graviton anomalous
dimension can become large.  Here, we are particularily interested in
the asymptotic safety scenario for gravity
\cite{Weinberg,Niedermaier:2006ns,Percacci:2007sz,Litim:2008tt,Reuter:1996cp,Souma:1999at,Lauscher:2001ya,Percacci:2002ie,Litim:2003vp,Fischer:2006fz,Codello:2007bd,Benedetti:2009gn,Niedermaier:2009zz}. We
also evaluate asymptotic freedom in the scenario where quantum gravity
is shielded by black hole formation
\cite{Banks:1999gd,Giddings:2001bu,DvaliGomez,Dvali:2010ue,Percacci:2010af}.  \step

The outline of this Letter is as follows. We introduce our renormalisation group set-up (Sec.~\ref{RG}), and discuss gravitational corrections to Yang-Mills theory within a background field approach (Sec.~\ref{AF}). A detailed discussion of the background field dependence is given (Sec.~\ref{BFD}), amended by a complementary RG study for flat backgrounds (Sec.~\ref{FB}). Our results are compared with earlier findings in perturbation theory at the one-loop order (Sec.~\ref{PT}). Beyond one-loop, we study the leading gravitational corrections to Yang-Mills in the presence of a gravitational fixed point (Sec.~\ref{GFP}) and in the presence of a cosmological constant term (Sec.~\ref{CC}). We close with a brief discussion   (Sec.~\ref{sec:disc}). \step

\section{Renormalisation group}\label{RG} 

The Functional Renormalisation Group is based on a momentum cutoff for
the propagating degrees of freedom and describes the change of the
scale-dependent effective action $\Gamma_k$ under an infinitesimal
change of the cutoff scale $k$
\cite{Wetterich:1992yh,Litim:1998nf,Gies:2006wv,Pawlowski:2005xe}.
Thereby it interpolates between a microscopic action in the
ultraviolet and the full quantum effective action in the infrared,
where the cutoff is removed. In its modern formulation, the
renormalisation group flow of $\Gamma_k$ with the logarithmic scale
parameter $t=\ln k$ is given by Wetterich's flow equation
\cite{Wetterich:1992yh}
\begin{equation}\label{eq:FRG}
\partial_t\Gamma_k=
\012\Tr\0{1}{\Gamma_k^{(2)}+R_k}\,\partial_t R_k\,.
\end{equation}
The trace stands for a sum over indices and a loop integration, and
$R_k$ (not to be confused with the Ricci scalar $R(g_{\mu\nu})$) is an
appropriately defined momentum cutoff at the momentum scale
$q^2\approx k^2$. The effective action depends on the metric field
$g$, the gauge field $A_\mu^a$ including possible Abelian factors, and
the related ghosts $\eta,\bar \eta$ and $C,\bar C$ for metric field
and gauge field respectively. In short we have $\Gamma_k=\Gamma_k
[\phi,\bar\phi]$, where the fields are put into a superfield $\phi=
(g,\eta,\bar\eta,A,C,\bar C)$. The action may also depend on
specifically chosen, non-propagating background fields $\bar \phi$. In
quantum gravity , the introduction of a background metric is necessary
as one has to fix a gauge, or more generally, one has to choose a
specific parameterisation of the configuration space. This fact is at
the root of the problem of background independence of quantum gravity.
\step

The flow equation \eq{eq:FRG} is derived from the insertion of infrared
cut-off terms in the path integral,
\begin{equation}\label{eq:Sk}
  \Delta S_k[\phi,\bar\phi]=\012 \int d^d x\sqrt{\bar g}\,
  (\phi-\bar \phi)_i\, R_k^{ij}[\bar\phi]\, (\phi-\bar \phi)_j\,,
\end{equation} 
where the regulator $R_k$ is a matrix in field space and may also
couple different species of fields, in particular metric and gauge
fields, and may depend on the background $\bar \phi$. Note that the
background metric dependence in \eq{eq:Sk} via $\sqrt{\bar g}$ is
necessary to keep the one loop nature of the flow equation
\eq{eq:FRG}.
Furthermore the 
contraction of the spin one and spin two components in $\phi$ 
also depend on the background metric. 
\step

The rhs.~of the flow equation \eq{eq:FRG} consists of one loop
diagrams with full field-dependent propagators multiplied with the
scale derivative of the regulator. This entails that for computing a
$n$-loop effective action one has to insert the $n-1$-loop effective
action on the rhs.  If interested in the one loop effective action we
therefore have to insert the classical action on the rhs. The
structure of the flow equation is such that standard perturbation
theory is recovered to all loop orders by iteration
\cite{Litim:2002xm}.  \step

Gauge and diffeomorphism symmetry is encoded in the related
Slavnov-Taylor identities
\cite{Litim:1998nf,Freire:2000bq,Pawlowski:2005xe}.  These symmetry
identities are modified in the presence of the regulator term
\eq{eq:Sk}. In the vanishing cut-off limit, $k\to 0$, the
modifications also vanish and we are left with the standard symmetry
identities. Furthermore, there are combined transformations of both,
the background $\bar \phi$ and the fluctuation $\phi$, which leaves
the effective action invariant.  It is for that reason that
\begin{equation}\label{eq:Gammainv} 
\Gamma_k[\phi]=\Gamma_k[\phi,\phi]\, 
\end{equation}
 is both, gauge
 and diffeomorphism invariant. 
 \step

\section{Asymptotic freedom and gravitational interactions}\label{AF}

In the present work we consider an approximation to the effective
action which reproduces the full one loop effective action but allows
also to discuss properties beyond one loop. Such an approximation for
$\Gamma_k[\phi]$ defined in \eq{eq:Gammainv} is given by
\begin{widetext}
 \begin{equation}\label{eq:EHk}
\Gamma_k[\phi,\bar\phi]=
\int d^dx \sqrt{\det g_{\mu\nu}}
\left[\0{Z_{N,k}}{16\pi G_N}\left(-R(g_{\mu\nu})+2\bar\Lambda_k\right)
+\0{Z_{A,k}}{4 \gYM^2} F^a_{\mu\nu}F_a^{\mu\nu}\right]
+S_{\rm gh}+S_{\rm gf}+\Delta\Gamma_k\,.
\end{equation}
\end{widetext}
Here we have introduced the Einstein-Hilbert action and the classical
Yang-Mills action as well as gauge fixing and ghost contributions
$S_{\rm gh}$ and $S_{\rm gf}$. The gauge and diffeomorphism invariant
action, \eq{eq:Gammainv}, is defined as
$\Gamma_k[\phi]=\Gamma_k[\phi,\phi]$ where the gauge fixing term and a
part of $\Delta\Gamma_k$ vanishes. Due to quantum fluctuations, all
couplings have become running couplings and depend on the infrared RG-scale
$k$: the running Newton coupling
$G_{N,k}=G_{N}/Z_{N,k}$, the running cosmological constant $\Lambda_k$, 
and the running Yang-Mills coupling $\gYMk=\gYM^2/Z_{A,k}$. 
Further quantum corrections such as 
higher order curvature invariants and further matter-gravity
interaction terms can be generated as well and are contained in
$\Delta\Gamma_k$.  \step

In the present work we discuss the running of the fundamental
dimensionless couplings of the theory, the Yang-Mills coupling and the
dimensionless Newton coupling, 
\begin{equation}\label{eq:gN}
\gNk=\frac{k^2 G_N}{Z_{N,k}}\,.
\end{equation}
The running is encoded in the respective $\beta$-functions
($\beta_X\equiv k\partial_k X$) as 
\begin{eqnarray}\label{eq:betaS}
  \beta_{\gYM}=& \012 \eta_A\, \gYMk\,\qquad\qquad & {\rm with} \qquad
  \eta_A=-\0{\partial_t Z_{A,k}}{Z_{A,k}}\,,\\
\label{eq:betaN}
  \beta_{\gN}= &(2+\eta_N) \,\gNk\,\qquad 
  &{\rm with} \qquad  \eta_N=-\0{\partial_t Z_{N,k}}{Z_{N,k}}\,. 
\end{eqnarray} 
The above equations \eq{eq:betaS}, \eq{eq:betaN} reflect the fact that
in a background field approach the running couplings run inversely to
the wave-function renormalisations $Z$ of the respective fields. The
running of the $Z$'s splits into a pure Yang-Mills part (only
Yang-Mills loops/fluctuations) and a gravity part that also contains
internal graviton lines,
\begin{equation}\label{eq:split}
  \eta_A=\eta_{A,\rm YM}+\eta_{A,\rm grav}\,,\qquad \qquad \eta_N
  =\eta_{N,\rm grav}+\eta_{N,\rm YM}\,.
\end{equation}
In the present work we are specifically interested in the sign
$\eta_{A,\rm grav}$, since $\eta_{A,\rm YM}$ is proportional to the
Yang-Mills coupling and tends to zero in the UV if asymptotic freedom
is present. Hence, for $\eta_{A,\rm grav}<0$ at vanishing YM-coupling,
asymptotic freedom is supported by quantum gravity corrections,
whereas for $\eta_{A,\rm grav}>0$ it is spoiled. \step

So far we have not specified the regulator $R_k$. Indeed the full
effective action at $R_k\equiv 0$ (for $k=0$) cannot depend on this
choice, however at finite $k\neq 0$ with $R_k$ there will be a
$k$-dependence. In the following we shall exploit this freedom to
disentangle gravity and gauge theory at one loop.  To that end we use
a specific class of regulators which respects the renormalisation
group scalings of the underlying physical theory at $k=0$,
\begin{eqnarray} \label{eq:Rk} 
R_k[\bar\phi] =\Gamma_k^{(2)}[\bar\phi]\, r[\bar\phi]\,.
\end{eqnarray}   
The dimensionless regulator functions $r^{ij}$ are demanded to scale
trivially under the RG-scalings of the underlying theory, and
\begin{equation} \label{eq:Gamma2}
\Gamma_{k,ij}^{(2)}[\phi,\bar\phi]
= \0{\delta^2 \Gamma_k[\phi,\bar\phi]}{\delta\phi^i\delta\phi^j}
\end{equation} 
are the second derivatives of the effective action w.r.t.\ the full
fields $\phi$.  Regulators of the form \eq{eq:Rk} are called
RG-adjusted \cite{Pawlowski:2001df,Pawlowski:2005xe} or spectrally
adjusted \cite{Gies:2002af}, and scale like (inverse) two-point
functions. With this property the RG-invariance and scaling properties
of the underlying theory carry over to that of the regularised theory
\cite{Pawlowski:2005xe}, and facilitate the study and interpretation
of $\beta$-functions w.r.t.\ to the $t$-scaling.  The dimensionless
regulator functions $r^{ij}$ depend on covariant Laplacians in the
presence of the background $\bar\phi$. We specify
\begin{eqnarray}\label{eq:r} 
  r^{gg}&=& r^{gg}(-\Delta_{\bar g} /k^2)
  \nonumber\\
     r^{\bar\eta\eta}&=&- r^{\eta\bar\eta}=r^{\bar\eta\eta}(-\Delta_{\bar g} /k^2)
     \nonumber \\
  r^{AA} &=&r^{AA}(-\Delta_{\bar g}(A)/k^2)\nonumber \\
  r^{\bar C C}&=&- r^{C\bar C}=r^{\bar C C}(-\Delta_{\bar g}(A) /k^2)\,. 
   \end{eqnarray} 
   All other components vanish. Regulator functions $r$ with \eq{eq:r}
   are diagonal. They only connect the metric and gauge field with
   themselves and gosts with anti-ghosts. Note that this is the
   minimal form of regulator functions that suppress all infrared (or
   ultraviolet) propagation of all fields. Dropping one of the
   components in \eq{eq:r} leads to full quantum propagation of the
   related field. Hence, the above choice naturally disentangles the
   Yang-Mills and gravitational sectors, and is therefore well-suited
   for our analysis.  \step

   Finally we are interested in the RG-scaling (w.r.t.\ $k$) of the
   couplings of the background gauge field $\bar A$ and the background
   metric field $\bar g$. Inserting the regulator \eq{eq:Rk} in the
   flow \eq{eq:FRG} and setting $\phi=\bar\phi$ we arrive at
   \cite{Pawlowski:2001df,Litim:2002xm,Litim:2002ce,Litim:2002hj,Pawlowski:2003sk,
     Pawlowski:2005xe,Gies:2002af}
\begin{equation}\label{eq:FRGadjusted}
\partial_t \Gamma_k[\phi]= \012 \Tr\, \0{1}{1+r[\phi]}\partial_t r[\phi]+
\Tr\, {\partial_t \Gamma_k^{(2)}[\phi,\phi]}\0{1}{\Gamma_k^{(2)}[\phi,\phi]}
\0{r[\phi]}{1+r[\phi]}\,,
\end{equation} 
where $\Gamma_k[\phi]=\Gamma_k[\phi,\phi]$, see \eq{eq:Gammainv}.  The
above RG flow is the key equation in the present work. The first term
on the rhs.~consists out of all one loop contributions to the flow.
The second term has the form of an RG improvement.  It consists out of
all terms beyond one loop, including all non-perturbative
contributions.  Note that it vanishes at one loop.  \step

We observe that for the cutoff choice with diagonal $r$ in \eq{eq:r},
the one loop flow disentangles.  It is the sum of the pure gravity
flow driven by $r^{gg}$ and $r^{\bar\eta\eta}$, and a pure gauge flow
driven by $r^{AA}$ and $r^{\bar C C}$.  Cross-terms leading to gravity
loops for i.e.~the gauge field propagators or gauge field loops for
the graviton propagators are absent.  The full one-loop effective
action is then given by
\begin{equation}\label{eq:oneloop}
\Gamma[\phi]=\Gamma_\Lambda[\phi]+\012 
\int_\Lambda^0 \0{dk}{k}\Tr\, \0{1}{1+r[\phi]}\partial_t r[\phi]\,, 
\end{equation} 
where $\Gamma[\phi]=\Gamma_{k=0} [\phi]$. It follows from
\eq{eq:oneloop} that for the class of cutoff functions \eq{eq:r} the
full $\beta$-function of the background gauge coupling, $\beta_{\gYM}$,
in the presence of gravitational fluctuations is given by the standard
result without gravity 
\begin{eqnarray}\label{eq:nogravity} 
\beta_{\gYM}=\beta_{\gYM,\rm YM}=-\012 \eta_{A,\rm YM}
\end{eqnarray} 
up to corrections starting at the 2-loop level. Beyond one loop, the
potential structural disentanglement no longer holds true: $\partial_t
\Gamma_k^{(2)}/\Gamma_k^{(2)}$ is not diagonal and hence entangles
Yang-Mills and gravity sector.

\section{Background field dependence}\label{BFD}

The functional renormalisation group approach of Sec.~III allows us to classify 
and understand the scheme dependence of the $\beta$-functions of a general 
matter-gravity system. In particular this allows us to explain the scheme 
dependences observed so far for the Yang-Mills -- gravity theory, as well as to 
extract the physics stored in the $\beta$-functions, $i.e.$~the UV stability of the 
coupled system. We note in this context that the functional RG set-up also covers the standard 
perturbative regularisation schemes such as momentum cutoff, dimensional, Pauli-Villars, 
or heat kernel regularisation.\step

Scheme dependences of $\beta$-functions in the background field
approach have already been studied for general field theories
\cite{Litim:2002ce,Pawlowski:2005xe,Pawlowski:2001df,Pawlowski:2003sk}.
In \cite{Litim:2002ce} it has been shown that the YM-coupling receives
contributions from the background field dependence of the
regularisation.  These terms even can alter the one-loop
$\beta$-functions.  It has also been shown explicitly how to extract
the standard one-loop $\beta$-function with the help of an equation
which tracks the background field dependence of the regularisation.
In YM theory, the background field approach is a convenient choice.
In gravity, however, it becomes a necessity.  There, the background
field dependence of the regularisation is encoded in
\begin{equation}\label{eq:fullback}
  \int \0{1}{\sqrt{\bar g} }\0{\delta \sqrt{\bar g}R_k}{\delta \bar\phi} 
  \0{\delta \Gamma_k[\phi,\bar\phi]} 
  {\delta R_k}=  \012 \Tr\, \0{1}{\Gamma_k^{(2)}[\phi,\bar\phi]+
    R_k[\bar\phi]}
  \0{1}{\sqrt{\bar g} }\0{\delta R_k[\bar\phi]}{\delta \bar\phi}\,.
\end{equation} 
 It is
left to evaluate whether the background field dependence affects the
one-loop result \eq{eq:nogravity} for the background coupling as 
suggested by the above considerations. For the given cutoff functions 
\eq{eq:Rk}, \eq{eq:r}, we derive the regulator-induced background field 
dependence \eq{eq:fullback},
\begin{equation}\label{eq:back}
  \int  \0{1}{\sqrt{\bar g} }\0{\delta \sqrt{\bar g}R_k}{\delta \phi} 
  \0{\delta \Gamma_k[\phi,\phi]} 
  {\delta R_k}=  \012 \Tr\, \0{1}{1+r[\phi]}
  \0{\delta r}{\delta \phi}+\,
  \Tr\,  \0{1}{\sqrt{g} }\0{\delta \sqrt{\bar g}
    \Gamma_k^{(2)}}{\delta \bar\phi}[\phi,\phi]\0{1}
  {\Gamma_k^{(2)}[\phi,\phi]}
  \0{r[\phi]}{1+r[\phi]}\,,
\end{equation} 
already evaluated at $\phi=\bar\phi$. Note that the structure of
\eq{eq:fullback} resembles that of the flow equation
\eq{eq:FRGadjusted}. The first term is purely diagonal and one loop,
whereas the second term contains the non-perturbative contributions
and entangles gravity and Yang-Mills due to $(\delta
\Gamma_k^{(2)}/\delta \bar \phi)/\Gamma_k^{(2)}$. In contradistinction
to the second term in \eq{eq:FRGadjusted}, it also contributes at one
loop. In particular, contributions proportional to $\delta
\Gamma_k^{(2)}/\delta \bar \phi$ in the gravity sector produce terms
\begin{equation}\label{eq:bargterms}
\propto\frac{\delta}{\delta\phi} \int d^4 x \sqrt{g}\, \tr\, F_{\mu\nu} F^{\mu\nu}\,,
\end{equation}
if evaluated at $\phi=\bar\phi$. Therefore we conclude that in general
graviton contributions to the regularisation-induced background
dependence of the effective action are created. We have already argued
that for the regularisation schemes diagonal in $r$ defined in
\eq{eq:r}, the graviton-contribution to the $\beta$-function of the background field defined
via \eq{eq:FRGadjusted} vanishes. Note that this does not hold for the
$\beta$-functions of the fluctuations. However, the $\beta$-function
of the background field also contains unphysical contributions from
the field-dependence of the regulators.  These terms are given by the
$t$-derivative of the $F^2$-contribution of the right hand side in
\eq{eq:back}. While the $t$-derivative of the $F^2$-contribution of
the first term vanishes, this is not so for the one-loop contribution
of the second term.

 In turn, regularisations which
induce a mixing between gravity and Yang-Mills fields are expected to
give a non-vanishing contribution already at one-loop for all
$\beta$-functions. 
The above arguments entail that the background $\beta$-functions of the coupled 
YM-gravity system are scheme-dependent even at one-loop.
Note, however, that the UV stability of the YM-gravity
system is solely controlled by the signs of the $\beta$-functions of
the fluctuation fields.  \step

In conclusion, the following picture has emerged. UV stability of the YM sector relates 
to the (negative) sign of the $\beta$-function for the fluctuating gluon. This information 
can be extracted directly from the correlation functions of the fluctuating gluons. 
Alternatively, one can exploit the fact that the regularisation-independent part 
of the $\beta$-function of the background coupling carries the same sign. In general, 
the second option requires the use of \eq{eq:back}.

\section{Flat backgrounds}\label{FB}

The results of the last section entail that a computation
of the graviton contribution to the Yang-Mills $\beta$-function
requires the distinction between fluctating graviton and background
graviton. This is possible if the computation is done in a flat
background within a standard vertex expansion. Here we put forward a
computation of the YM $\beta$-function, which differs from the
previous studies in two aspects \cite{Folkerts:Dipl2009}: Firstly, the
analysis is performed for trivial backgrounds $\bar A=0$ and $\bar
g_{\mu\nu}=\eta_{\mu\nu}$. Secondly, we employ classes of momentum
cutoffs with a tensorial structure different from \eq{eq:Rk}. The
result complements our previous findings and allows to make close
contact with all previous 1-loop studies
\cite{Ebert:2007gf,Toms:2008dq,Toms:2007sk,Robinson:2005fj} as well as
the flow study \cite{Daum:2009dn}.\step

This also allows us to make use of the fact that loop contributions of
fluctuating gravitons to the correlation functions of the fluctuating
gluons and to the background gluons agree: the vertices of the two
fields are derived only from the classical YM-action and not from the
ghost and gauge fixing terms. The classical YM-action only depends on
$\phi= \bar\phi+(\phi-\bar\phi)$ and hence derivatives w.r.t.\ the
background fields $\bar\phi$ and w.r.t. the fluctuation fields
$(\phi-\bar\phi)$ agree. In other words, the two possibilities of how
to compute the graviton contribution to the $\beta$-function of the
fluctuating gluon agree. We emphasise that this does not hold for the
gluon contribution to the $\beta$-function of the fluctuating
graviton.\step

Our Ansatz for the effective action is dictated by asymptotic
freedom: a good approximation for Yang-Mills theory at
high energies is given by the classical $F^2$ operator plus a gauge
fixing term, both equipped with a running wave function
renormalisation $\ZA$ and a classical scale-independent ghost
action. The interaction of the gauge bosons with gravity is
induced by the metric appearing in the spacetime integral measure
which is now a quantum field. Also adding the Einstein-Hilbert
action, we arrive at
\begin{eqnarray}
\label{truncation for ymwgravity}   
\Gamma_k[g,A;\bar{g},\bar{A}]
&=&\frac{Z_{N}}{16\pi{G}_N}\int d^dx\sqrt{g}\left(-R+2\bar{\lambda}_k\right)
+\frac{Z_{A}}{4}\int d^dx\sqrt{g}g^{\mu\rho}g^{\nu\sigma}F_{\mu\nu}^aF_{\rho\sigma}^a
\nonumber\\
&+&\frac{Z_{N}}{2\alpha}\int d^dx\sqrt{\bar{g}}\bar{g}^{\mu\nu}L_\mu L_\nu\
+\frac{Z_{A}}{2\xi}\int d^dx\sqrt{\bar{g}}(\bar{g}^{\mu\nu}
\bar{D}^{a b}_\mu (A-\bar{A})^b_\nu)^2+S_{\mathrm{gh}}\,.\quad
\end{eqnarray}
Note that the $k$-dependence of $Z_N$ and $Z_A$ is not indicated
explicitly.  In \eq{truncation for ymwgravity}, we have a general
linear gauge condition $L_\mu$ with gauge parameter $\alpha$ for the
graviton, and a background field gauge with parameter $\xi$ for the
gluon. In general, both gauge fixing parameters are $k$-dependent,
except for vanishing values. The latter constitutes a fixed point of
the flow, \cite{Litim:1998qi}. $S_\mathrm{gh}$ consists of the sum of
the classical ghost action for gravity and the one for Yang-Mills
theory.  $S_\mathrm{gh}$ neither contributes to the graviton-induced
corrections to the running of the gluon coupling, nor to the
gluon-induced running of the gravitational coupling.  There is a
simple reason for this. The gravitational ghost term, by construction,
does not contain any gauge fields. The Yang-Mills ghost term only
couples to the background metric field, eg.~$S_{\rm YM, gh}=\int
d^dx\sqrt{\bar{g}}\,\bar{g}^{\mu\nu}\,\bar{c}^a\,\bar{D}^{a b}_\mu\,
D^{b c}_\nu\, c^c$ in a covariant gauge. Since $\gammatwo$ is obtained
by differentiation with respect to the dynamical fields, the gauge
fixing terms do not generate interaction terms between gravity and the
gauge field ghosts. \step

Hence the key input in the flow equation is the full propagator,
$1/(\gammatwo_k+R_k)$ for the gauge field $A$ and the metric field
$g$. We introduce
\begin{equation}
	\label{matrices for gamma2 and rk}
	\gammatwo_k=
	\begin{pmatrix} 
	\gammatwo_{AA}&&\gammatwo_{Ag}\\ \\ 
\gammatwo_{gA}&&\gammatwo_{gg}\end{pmatrix}
	\qquad\mathrm{and}\qquad R_k=
        \begin{pmatrix} R_{A}&0\\ 0&R_{g}\end{pmatrix},
\end{equation}
with the convention $\gammatwo_{\varphi_i\varphi_j}\equiv
\frac{\delta^2\Gamma_k}{\delta\varphi_i\delta\varphi_j}$ for
$\varphi=\{A,g\}$. Note that we have chosen a regulator $R_k$ which is
diagonal. An important observation is that the off-diagonal terms of
$\gammatwo$ in \eq{matrices for gamma2 and rk} vanish for
$A=\bar{A}=0$ and $g=\bar{g}=\eta$ since the gauge field part in our
ansatz involves at least two gauge fields.  Consequently, the inverted
matrix reads
\begin{equation}
	\label{prop zero fields}
\displaystyle	\frac{1}{\Gamma^{(2)}_k+R_k}\Ofield=\begin{pmatrix} 
\displaystyle	\frac{1}{\gammatwo_{AA}+R_{AA}}&&0\\\\ 
          0&&
\displaystyle          \frac{1}{\gammatwo_{gg}+R_{g}}\end{pmatrix}.
\end{equation}
We specify the
regulator functions and their tensor structures,
\begin{eqnarray}
	\label{general form of regulators}
	R_{A}&=&Z_A\;{T}_A(p)\; p^2\; r_A(p^2/k^2)\,,\\
	\label{Rgrav}
	R_{g}&=&\frac{Z_N}{32 \pi G_N}\;\frac{1}{2}\,{T}_g(p)\; p^2\; r_g(p^2/k^2)\,.
\end{eqnarray}
Here $r(p^2/k^2)$ denotes the shape for the scalar part of the
momentum cutoff, which can range between $0\leq r(y)\leq \infty$ for
$\infty\ge y\ge 0$. Both regulators are dressed with the appropriate
wave function renormalisation factor to ensure that $R$ displays the
same RG scaling as $\Gamma^{(2)}$. The tensor structures ${T}_A$ and
${T}_g$ are chosen as
\begin{eqnarray}
	\label{cutoff gluon}
	T^A_{\mu\nu}\hspace{-1mm}&=&\hspace{-1mm}
	\eta_{\mu\nu}-(1-\frac{1}{\xi})\frac{p_\mu p_\nu}{p^2}\,,
		\\
	\label{cutoff grav}
        T^g_{(\alpha\beta)(\gamma\delta)}\hspace{-1mm}&=&\hspace{-1mm}
        \frac{\alpha-1}{\alpha}\left[
          2\left(\eta_{\gamma\delta}\frac{p_\alpha p_\beta}{p^2}
            +\eta_{\alpha\beta}\frac{p_\gamma p_\delta}{p^2}\right)
          -\eta_{\beta\gamma}\frac{p_\alpha p_\delta}{p^2}-\eta_{\beta\delta}
          \frac{p_\alpha p_\gamma}{p^2} -\eta_{\alpha\gamma}
          \frac{p_\beta p_\delta}{p^2}-\eta_{\alpha\delta}
          \frac{p_\beta p_\gamma}{p^2}\right]\nonumber\\
        &&
        +\left[
          \eta_{\alpha\gamma}\eta_{\beta\delta}+\eta_{\alpha\delta}\eta_{\beta\gamma}
          +\frac{1-2\alpha}{\alpha}\eta_{\alpha\beta}\eta_{\gamma\delta}\right]\,,
\end{eqnarray}
with
$T^g_{(\alpha\beta)(\gamma\delta)}=T^g_{(\gamma\delta)(\alpha\beta)}$,
and $T_A$ diagonal in colour space. We extract the graviton
contributions to the running for the YM coupling from the graviton
contributions to the running of the YM propagator.  The relevant diagrams on the rhs of the flow equation are
\begin{eqnarray}
\label{RHS gluon flow}
\begin{minipage}[c]{2.5cm}
\includegraphics[width=2.5cm]{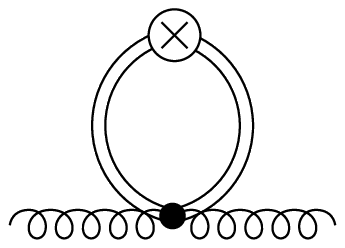}
\end{minipage}+
\begin{minipage}[c]{3.6cm}
\includegraphics[width=3.6cm]{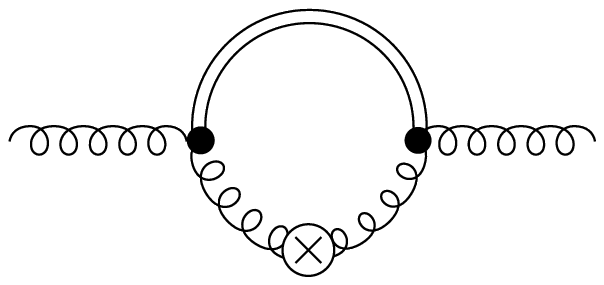}
\end{minipage}
&&\!\!\!\!
+
\begin{minipage}[c]{3.6cm}
\includegraphics[width=3.6cm]{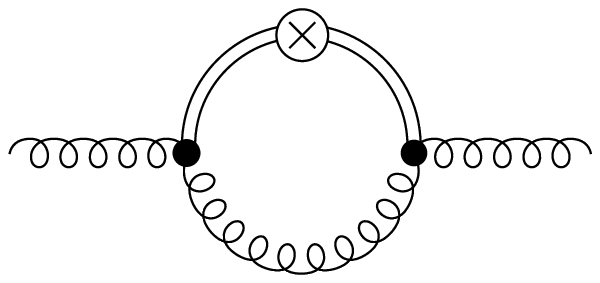}
\end{minipage}
\end{eqnarray}
where double lines denote the full graviton propagator, curly
lines the full gluon propagator, full dots the full
vertices, and the crossed circle the appropriate regulator insertion
$\partial_t R_k$.\step

Now we come to an important observation. There is a kinematical
identity which links the tadpole diagram, the first term in \eq{RHS
  gluon flow}, to the other two self energy terms in \eq{RHS gluon
  flow}.  This identity is expressed diagramatically in
\eq{eq:angularaverage}: It states that the contributions to the tree level gluon-gluon-graviton-graviton amplitude from the 4-point vertex and from the 3-point vertices are proportional to each other when averaged over their angular dependencies.
This holds for
arbitrary gauge fixing parameter $\alpha$, and in the presence of a
cosmological constant.  This result can be written as
\begin{equation}
\label{eq:angularaverage}
\includegraphics[width=0.55\textwidth]{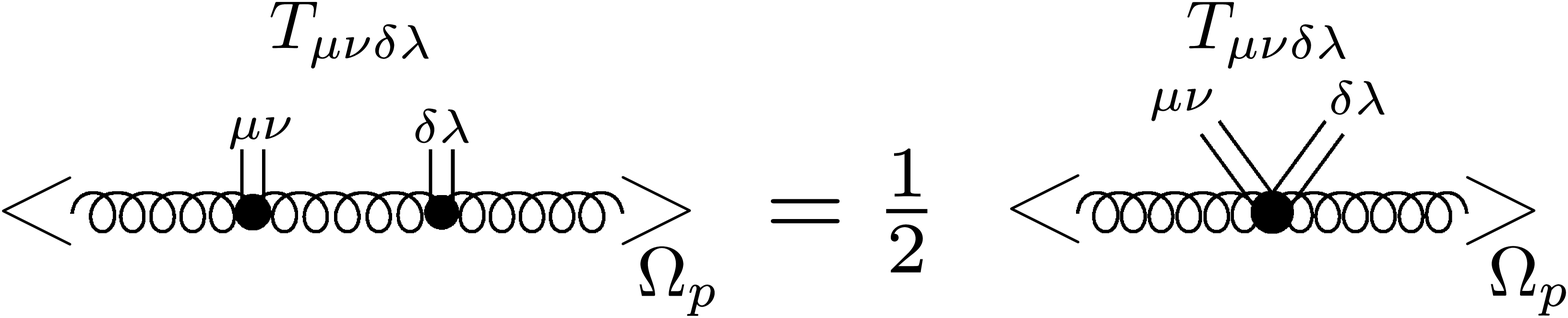}
\end{equation}
and holds for contractions of the open-ended graviton legs with any of
the tensor structures $T_{\mu\nu\delta\lambda}$ of the graviton
propagator itself. The brackets indicate that this equality is only
valid after the integration over $\int d\Omega_p$.  \step

The relation \eq{eq:angularaverage} relates to the fact that the
graviton-gluon vertices are derived from the classical kinetic term of
Yang-Mills theory. This approximation to the effective action of
Yang-Mills certainly holds at the Gau\ss ian, asymptotically free
fixed point of Yang-Mills. We conclude that it should also hold in the
presence of gravity if the gravity contributions to Yang-Mills sustain
asymptotic freedom. We shall use this selfconsistency argument later.\step

A general regularisation scheme may break explicitly or implicitly the
relation \eq{eq:angularaverage} as it modifies either the propagation
or the dynamics of the theory, or both, by cut-off effects. For
example, a regularisation scheme in momentum space only changes the
propagator, whereas a cutoff in covariant momenta changes propagators
and vertices, and might require even tree-level counter terms. In
turn, a symmetry preserving regularisation maintains the symmetry
relation \eq{eq:angularaverage} at the quantum level. \step

In order to extract the flow equation for the gluon wavefunction
renormalisation $\partial_t {Z}_{A}$ we project both sides of the flow
equation onto the transversal part of the inverse gluon propagator
with $\Pi_{\mu\nu}^T=(\eta_{\mu\nu}-\frac{q_\mu
  q_\nu}{q^2})$. Equating the coefficients in front of $q^2$ we
identify the gravitational contribution to the flow for the gluon wave
function renormalisation as
\begin{equation}
\label{eq:ZAlambda0}
\frac{\partial_t Z_A}{Z_A} =48 \pi^3 \frac{G_N }{Z_N}\int_0^\infty\frac{d p\:p}{(2\pi)^4}
\frac{(1+\alpha)}{(1+r_{g})}\left[\frac{
    \partial_t(Z_A r_{A})}{Z_A(1+r_{A})^2}+\frac{\partial_t(Z_N r_{g})}{
    Z_N(1+r_{A})(1+r_{g})}-\frac{\partial_t(Z_N r_{g})}{Z_N(1+r_{g})}\right].
\end{equation}
In \eq{eq:ZAlambda0} we have dropped terms proportional to the flow of
the gravitiational gauge fixing parameter, $\partial_t\alpha$, its
impact will be discussed later. Note also that there are no terms
proportional to $\xi$ and $\partial_t \xi$ due to transversal projections
related to the tensor structure of the graviton -gluon vertices. With
the anomalous dimensions $\eta_A$ and $\eta_N$, see \eq{eq:betaS} and
\eq{eq:betaN} respectively, we deduce from \eq{eq:ZAlambda0} that
\begin{equation}
\label{eq:etaA0}
\eta_{A,\rm grav}= -\frac{3\,I}{\pi}\gNk \quad{\rm with}\quad I=I_0-\frac{\eta_A}{2}\,I_1\,.
\end{equation} 
with the dimensionless running Newton coupling defined in \eq{eq:gN}. The coefficients
\begin{eqnarray}
\label{eq:IetaAgrav}
  I_0&=&\int_0^\infty dx\;\frac{1+\alpha}{1+r_g}\0{ r_A}{1+r_A}
  \left(1+\012 \0{ \eta_N r_g}{
        1+r_g}\right) \\
\label{eq:IetaAgrav1}
          I_1&=&\int_0^\infty dx\;\frac{1+\alpha}{1+r_g}\0{ r_A}{(1+r_A)^2}
\end{eqnarray}
originate from the radial momentum integration with $x=p^2/k^2$. As
discussed before, they only depend on the gravitational gauge fixing
parameter but not on the Yang-Mills gauge fixing parameter. The
dropped term proportional to $\partial_t\alpha$ would effectively
lower the coefficient $1/2$ in front of the $\eta_N$-term in $I_0$. To
see this we remark that in leading order the gauge fixing term is
$k$-independent. This entails that the gauge fixing parameter runs
like the wave function renormalisation, $\partial_t\ln \alpha\propto
-\eta_N$. As the prefactor of the $\eta_N$-term comes from a summation
over all tensor component, the running of $\alpha$ effectively removes
the gauge fixing direction from this sum and hence lowers the
prefactor. This structure is also present beyond leading order,
indeed, for $\alpha=0$ the related term even vanishes
identically. Thus, for the sake of simplicity we shall drop this term
as it does not change the arguments presented here.  Solving
\eq{eq:etaA0} for $\eta_{A,\rm grav}$ we arrive at
\begin{equation}\label{eq:etaA}
  { \eta_A} =\eta_{A,{\rm YM}}+\eta_{A,{\rm grav}}=\0{\eta_{A,{\rm \tiny YM}}- 
    \s03\pi {\gNk}I_0}{1-\s0{3}{2\pi} {\gNk}I_1} \,. 
\end{equation}
We conclude from \eq{eq:etaA} with the coefficients \eq{eq:IetaAgrav},
\eq{eq:IetaAgrav1} that the gravity-induced running of Yang-Mills is
not universal. Note however that the regulators in \eq{eq:IetaAgrav}
and \eq{eq:IetaAgrav1} in general, do not respect the symmetry
constraint \eq{eq:angularaverage}.

\section{Perturbation theory} \label{PT}

The one-loop perturbative results are recovered from the flow equation
by using the bare two-point functions on the right-hand side of the
flow. This corresponds to setting $Z_A=Z_N=1$ on the right hand side
of \eq{eq:etaA}. Then the one-loop graviton correction to the gluon
anomalous dimension reads
\begin{equation}
\label{1loop lambda0}
\eta_{A,\rm grav}=-\frac{3 \gN}{\pi}\,I_{0,\rm 1-loop}\,,
\end{equation}
with
\begin{equation}\label{I}
I_{0,\rm 1-loop}=\int_0^\infty dx\;\frac{1+\alpha}{1+r_g(x)} \frac{r_A(x)}{1+r_A(x)}\,.
\end{equation}
$I_{0,\rm 1-loop}$ encodes the gauge-fixing and regularisation dependence of the
one-loop coefficient. It is crucial to observe that
\begin{equation}
	\label{eq:eta_Agrav}
	\eta_{A,\rm grav}\Big|_{\rm 1-loop}\leq0\,,
\end{equation}
for $\alpha\geq -1$, and all regulators. 
We conclude that the full one-loop running of Yang-Mills is not universal
due to the gravity correction. However, it sustains asymptotic freedom
for all regularisations. Furthermore it has both, a regulator as well
as a gravity gauge dependence via $\alpha$. Both are not independent
as the latter one can be absorved within a specific choice of the
regulator. This is not surprising as the regulator can be also
partially viewed as a change of the gauge fixing. In order to
disentangle these effects we first consider a vanishing gravity
regularisation, $r_g\equiv 0$. The gauge dependence persists,
$\partial_\alpha \eta_{A,\rm grav}=\eta_{A,\rm grav}/(1+\alpha)$ at
fixed Newton coupling.  Hence, $\eta_A$ is only independent of
$\alpha$ for $\eta_{A,\rm grav}\equiv 0$, which is equivalent to the
constraint
\begin{equation}\label{eq:zerogen} 
\0{1}{1+r_g}\frac{r_A}{1+r_A}\equiv 0\,. 
\end{equation} 
The constraint \eq{eq:zerogen} implies that $1/r_g$ has to vanish for finite $r_A>0$.
Consequently, $r_g$ is a sharp cutoff for momenta where $r_A>0$. 
Legitimate choices are for example $r_g=r_A=r_{\rm sharp}$, or
$r_g=r_{\rm sharp}$ and $r_A=r_{\rm opt}$ \cite{Litim:2000ci}. The sharp and optimised regulators 
are defined as 
\begin{eqnarray}\label{eq:sharp} 
  r_{\rm sharp}(x)&=&\0{1}{\theta(x)}-1\,, \\
\label{eq:opt} 
r_{\rm opt} (x)&=&\left(\0{1}{x}-1\right) \theta(1-x)\,,
\end{eqnarray}
and the Heaviside step function $\theta$.  Note that regulators with
the constraint \eq{eq:zerogen} effectively satisfy the symmetry constraint
\eq{eq:angularaverage}: if the graviton legs in \eq{eq:angularaverage}
are contracted with graviton propagators, we are lead to
\eq{eq:zerogen}. In turn, without graviton propagators the constraint
\eq{eq:angularaverage} necessitates $r_A=0$.\step

We summarise the above analysis as follows: we have computed the one
loop gravity contributions $\eta_{A,\rm grav}$ to the Yang-Mills
$\beta$-function for general regularisation schemes. 
The one-loop gravitational contribution to the running of the YM
coupling is regularisation-dependent, and thus non-universal. This
originates from the mass dimension inherent to the gravitational
coupling. However, $\beta_{\textrm{\tiny YM},\rm grav}$ is negative
semi-definite and hence sustains asymptotic freedom.  It vanishes in
symmetry-preserving regularisation schemes based on the symmetry
relation \eq{eq:angularaverage}.

\section{Asymptotic freedom and quantum gravity}\label{GFP}

In this section we extend the stability analysis of asymptotic freedom
to include general field-theory-based gravity scenarios
\cite{Weinberg,Niedermaier:2006ns,Percacci:2007sz,Litim:2008tt,%
  Niedermaier:2009zz,Reuter:1996cp,Souma:1999at,Lauscher:2001ya,%
  Percacci:2002ie,Litim:2003vp,Fischer:2006fz,%
  Codello:2007bd,Benedetti:2009gn,%
  Banks:1999gd,Giddings:2001bu,DvaliGomez,Dvali:2010ue}. In some of
these, short distance physics is shielded by black hole formation
\cite{Banks:1999gd,Giddings:2001bu,DvaliGomez,Dvali:2010ue}. Provided
this happens at scales where the anomalous dimension of the graviton
$\eta_N$ is still small, the weak coupling analysis of the previous
section is sufficient to sustain asymptotic freedom of Yang
Mills. However, since the graviton anomalous dimension $\eta_N$ might
grow large due to strong quantum gravity effects, it is important to
evaluate the stability of asymptotic freedom for general $\eta_N$.
\step

From the RG-equation for $\gN$, \eq{eq:betaN}, we note that the value
$\eta_N=-2$ is distinguished because the $\beta$-function for the
dimensionless Newton coupling $\gN$ vanishes. This is the
gravitational fixed point of the asymptotic safety scenario for
gravity
\cite{Weinberg,Niedermaier:2006ns,Percacci:2007sz,Litim:2008tt,%
  Niedermaier:2009zz,Reuter:1996cp,Souma:1999at,Lauscher:2001ya,%
  Percacci:2002ie,Litim:2003vp,Fischer:2006fz,%
  Codello:2007bd,Benedetti:2009gn}.  In its vicinity, the
gravitational $\beta$-function \eq{eq:betaN} changes sign. Therefore
the three cases $\eta_N=-2$, $\eta_N<-2$ and $\eta_N>-2$ are
qualitatively different and will be discussed separately. \step

We first discuss the impact of a gravitational fixed point $\eta_N=-2$ 
for asymptotic 
freedom of Yang-Mills by evaluating the gravitational
contribution $\eta_{A, \rm grav}$, \eq{eq:etaA0}. The related 
coefficient $I_0$ in \eq{eq:IetaAgrav} becomes
\begin{equation}\label{eq:etaAgravfull}
  I_0=\int_0^\infty dx\;\frac{1+\alpha}{(1+r_g)^2} \0{r_A}{(1+r_A)}
\end{equation}
and the coefficient $I_1$ is given by \eq{eq:IetaAgrav1}. With the
coefficients $I_0$ and $I_1$ and the assumption of asymptotic freedom,
$g_{\rm YM}\equiv 0$ at the fixed point, the $\beta$-function at the
fixed point \eq{eq:etaA} reads
\begin{equation}\label{eq:etafixedA}
 { \eta_A}_* =- \0{\s03{\pi}  {g_N}_* I_0}{1-\s0{3 }{2\pi} {g_N}_* I_1 } \,. 
\end{equation}
As for the one loop running, ${\eta_A}_*$ is not universal since the
coefficients $I_0,I_1$ depend on the choices for the regulators
$r_g,r_A$. Asymptotic freedom only enforces ${\eta_A}_*\leq 0$ which
implies
\begin{equation}\label{eq:constraints} 
  I_0\geq 0 \quad \wedge\quad  {g_N}_*\leq \frac{2 \pi}{3 I_1}\, ,
  \qquad {\rm or}\qquad I_0< 0 \quad \wedge \quad {g_N}_* > \frac{2 \pi}{3 I_1}\,.
\end{equation}
The latter case is irrelevant here as $I_0< 0$ can only be obtained
for the singular choices $\alpha<-1$. For the symmetry preserving
choices \eq{eq:zerogen} we have $I_0=0$ and we arrive at a vanishing
non-perturbative gravity contribution for the Yang-Mills
$\beta$-function, ${\eta_A}_*=0$.  \step

In the general case we have to satisfy the constraint in
\eq{eq:constraints} relating $I_1$ and ${g_N}_*$. For the sake of
simplicity we restrict ourselves to regulators with $r_A(x\geq
1)\equiv 0$ as this limits the size of the back-reaction of the gluon
on the graviton. Then, $I_1$ in \eq{eq:IetaAgrav1} satisfies
\begin{equation}\label{eq:etaAgravbound2}
  I_1=\int_0^1 dx\;\frac{1+\alpha}{1
    +r_g} \0{r_A}{(1+r_A)^2}\leq \0{1}{4}
  \frac{1 +\alpha}{1+r_{g,\rm min}}\,, 
\end{equation}
where $r_{g,\rm min}=\min r_g(x)$ with $x\in [0,1]$. We have also used
that $r_A/(1+r_A)^2\leq 1/4$. Due to the bound \eq{eq:etaAgravbound2}
the first set of constraints in \eq{eq:constraints} is satisfied for
\begin{equation}\label{bound}
{g_N}_*\le
\frac{8\pi}{3}\frac{1+\alpha}{1+r_{g,\rm min}}\,.
\end{equation}
The bound \eq{bound} for ${g_N}_*$ is safely satisfied for general
regulators $r_g$.  Indeed, the generic value for the fixed point
coupling is of order one. For example, for the pure gravity system one
finds ${g_N}_*=0.893$ for an optimised regulator and $\alpha=0$.  Note
also in this context that ${g_N}_* (1+\alpha)$ is approximately
constant for all $\alpha$. If we relax the locality condition for the
gluon, that is $r_A(x\geq 1)\not\equiv 0$, one enhances the gluon
fluctuations. In this case one has to take into account the
back-reaction of the gluon fluctuations on the fixed point value of
the Newton coupling. Here we only remark that in the full system
indeed one can show that $ {g_N}_*\leq \frac{2 \pi}{3 I_1}$ is always
satisfied. This will be discussed in more detail elsewhere. 

In summary we find that the graviton contribution
$\eta_{A,\rm grav}$ sustains asymptotic freedom
\begin{equation}
	\label{lambda0 result}
{\eta_{A,\rm grav}}_*\leq 0\,. 
\end{equation} 
Hence the ultraviolet fixed point action of Yang-Mills theory is given by the
classical action and we can invoke \eq{eq:angularaverage} leading to
$\eta_{A,\rm grav}=0$. This leads us to the final value of the
non-perturbative $\beta$-function if all symmetries are respected,
\begin{equation}\label{eq:finalbeta} 
{\eta_{A}}_*=0\,. 
\end{equation}
We conclude that the gauge coupling remains asymptotically free, even
in the presence of a non-perturbative gravitational fixed point.  \step

Next we turn to $\eta_N<-2$. In this case $I_0$ might turn negative, see
\eq{eq:IetaAgrav}. However, with \eq{eq:betaN} we have $\beta_{\gN}<0$ and
$g_N$ decreases exponentially with logarithmic RG scale $t=\ln k$, 
and so does the gravity contribution to $\eta_A$ in
\eq{eq:etaA}. We conclude that even though the gravity contribution to
$\eta_A$ might be positive for $\eta_N<-2$, its exponential decay
would eventually lead to the dominance of $\eta_{A,YM}$. Thus,
asymptotic freedom would persist. Note also that $\eta_N<-2$ for
$t\to\infty$ is equivalent with a Gau\ss ian UV fixed point which is
not present in gravity. \step

Finally, for $\eta_N>-2$ the dimensionless Newton coupling 
$g_N$ grows exponentially with
$t$, see \eq{eq:betaN}, and the coefficient $I_0$ is positive. This
case includes classical gravity with $\eta_N=0$ as well as a
gravitational shielding scenario. The constraints \eq{eq:constraints}
enforce $g_N\leq 2 \pi/(3 I_1)$  to ensure asymptotic freedom for 
Yang-Mills. This puts an upper limit on the value of
dimensionless Newton coupling which has to be satisfied by  
eg.~perturbation theory or a gravitational shielding scenario, or else
$g_N$ de-stabilizes asymptotic freedom. Note that this bound depends 
on the regularisation scheme as does the definition of the dimensionless 
Newton constant $g_N$.\step

In summary we have shown that asymptotic freedom persists in general
field-theory-based gravity scenarios.

\section{Cosmological constant}\label{CC}
We proceed by introducing a cosmological constant $\Lambda$ to the
theory. For the sake of simplicity we consider the case $\alpha=0$,
but the results readily extend to $\alpha\neq 0$. The coefficient
functions $I_0$ and $I_1$, \eq{eq:etaA0}, read
\begin{eqnarray}\nonumber 
  I_0&=&\int_0^\infty dx\;\frac{1}{1- \0{2\lambda}{x}+ r_g} \0{r_A}{1+r_A}
  \left(1+\left(\0{\eta_N r_g}{2}-\0{2\lambda+\partial_t \lambda}{x}\right)\0{1}{
        1-\0{2\lambda}{x}+r_g}\right) 
        \\ 
  I_1&=&\int_0^\infty dx\;\frac{1}{1- \0{2\lambda}{x}+ r_g} \0{r_A}{(1+r_A)^2}
\label{eq:etaALambda}\end{eqnarray}
where $\lambda=\Lambda_k/k^2$ denotes the cosmological constant in
units of the RG scale $k$.  This leads to the beta-function
\eq{eq:etafixedA} with the coefficients $I_0$ and $I_1$ in
\eq{eq:etaALambda}.  As before we conclude that if invoking the
symmetry relation \eq{eq:angularaverage} leading to \eq{eq:zerogen}
with $1+r_g\to 1-2\lambda/x+r_g$, the coefficient function
\eq{eq:etaALambda} vanishes, $I\equiv 0$, and we have no gravity
contribution to the Yang-Mills running,
\begin{equation}\label{eq:finalbetaLambda} 
{\eta_{A,\rm grav}}=0\,. 
\end{equation}
It is left to prove the
stability of this result under changes of the regulator, that is
$\eta_{A,\rm grav}\leq 0$ for all regulators. 

As in the previous section we consider the qualitatively different
cases $\eta_N=-2$, $\eta_N<-2$ and $\eta_N>-2$.  At the gravitational
fixed point with $\eta_N=-2$ and $\partial_t \lambda=0$ we find
\begin{equation}\label{eq:>=}
  I_0=\int_0^\infty dx\; \0{r_A}{1+r_A} \,\frac{1}{(1-\frac{2\lambda}{x}+r_g)^2} 
  \left(1-\0{4\lambda}{x}\right) \,. 
\end{equation}
We have stability of the fixed point iff $I_0\ge 0$. Evidently the
integrand in \eq{eq:>=} is not positive but is negative for small
momenta $x$ and turns positive for $x >4 \lambda$.  For regulators
\eq{eq:sharp}, \eq{eq:opt} we find stability for
$\lambda\le\lambda_{\rm crit}=\frac{1}{8}$. For general regulators one
has to solve the integral equation $I_0$ numerically resulting in
$\lambda_{\rm crit}(r_g,r_A)$ and the constraint
$\lambda\le\lambda_{\rm crit}(r_g,r_A)$. The regulator-dependence of
the critical cosmological constant is not surprising as the value of
the cosmological constant (and the Newton constant) at the fixed point
is not a physical observable. The important result is the existence of
such a constraint.\step

For $\eta_N<-2$ and $\eta_N>-2$ we run into the same scenarios as in
the previous section with the coefficients $I_0$ and $I_1$. Note,
however, that for $\eta_N>-2$ we require $I_0>0$ which puts
constraints on $\lambda$ and $\partial_t \lambda$. As one can assume
$\partial_t \lambda<0$ due to the canonical running in this regime,
they are less severe than the one in the asymptotic safety scenario.

\section{Discussion}\label{sec:disc}

First we compare our results with previous findings to leading
order in perturbation theory in the $U(1)$ and $SU(N)$ cases
\cite{Deser:1974cz,Deser:1974nb,Robinson:2005fj,% 
Pietrykowski:2006xy,Toms:2007sk,Ebert:2007gf,%
Daum:2009dn,Toms:2010vy}.
Identifying the energy scale $E$ with the cutoff scale $k$, the
one-loop gravitational correction can be written as
\begin{equation}
	\label{1loop result}
	\beta_{\textrm{\tiny YM},\rm grav}=-\frac{3\, I}{2\pi}
        g_\textrm{\tiny YM}\,G_N\,E^2\,.
\end{equation}
Our result corresponds to the coefficient $I=I_{\rm 1-loop}$ as
given in \eq{I}. Within dimensional regularisation and using the
background field method,
Deser~et.~al.~\cite{Deser:1974cz,Deser:1974nb} found no gravitational
corrections to the Yang-Mills and Maxwell $\beta$-functions,
respectively, and hence $I=0$.  In the $U(1)$ case, the same
conclusion is reached by Pietrykowski \cite{Pietrykowski:2006xy} by
evaluating the gauge fixing dependence in a generalised $R_\xi$ gauge,
and by Toms \cite{Toms:2007sk} within the geometrical effective
action method.  In the $SU(N)$ case, Ebert, Plefka and Rodigast
\cite{Ebert:2007gf} confirmed $I=0$ based on dimensional
regularisation and cutoff regularisation with Feynman gauge for the
gluon and de Donder gauge for the graviton. This study shows that
quadratic divergences, neglected within dimensional regularisation,
cancel out within a sharp cutoff regularisation.\step

On the other hand, Robinson and Wilzcek \cite{Robinson:2005fj}, Toms
\cite{Toms:2010vy} and Daum et.~al.~\cite{Daum:2009dn} find a
non-vanishing contribution to leading order, $I>0$. In particular, the
result of Robinson and Wilczek \cite{Robinson:2005fj} is based on the
Feynman gauge $\xi=1$ and $\alpha=1$, leading to $I_{\rm RW}=2$ in
\eq{1loop result}. A positive coefficient $I_{\rm T}=\frac{2}{3}$ has
also been found by Toms \cite{Toms:2010vy} using a modified version of
the Vilkovisky de-Witt approach.  The study of Daum, Harst and Reuter
\cite{Daum:2009dn} is based on the background field method, together
with a Wilsonian momentum cutoff for the propagating modes, with
\begin{equation}\label{eq:DHR}
I_{\rm DHR}=4\int_0^\infty dx\frac{-x^2\, r'(x)}{1+r(x)}
=8\int_0^\infty dx\,x\, \ln [1+r(x)]\,.
\end{equation}
Here $r(x)$ parametrises the shape of the momentum cutoff. The
coefficient is non-universal and strictly positive, $I_{\rm DHR}> 0$,
as $r'\not\equiv 0$. As has been argued in Sec.~\ref{BFD}, the
background field approach used in this study does also include terms
originating in the background field dependence of the regulator.  For
the symmetry-preserving cut-off choice $r_{\rm sym}$ with
\eq{eq:zerogen} the coefficient \eq{eq:DHR} diverges in clear contradistinction 
to our result $I[r_{\rm  sym}]=0$. This difference
can be solely attributed to the field-dependence of the cut-off function which
enters \eq{eq:DHR}.\step

In the presence of a cosmological constant, a recent study based on a
diffeomorphism invariant expansion scheme using the geometric
effective action finds a non-vanishing result to one-loop order
\cite{Toms:2008dq}. The expansion scheme used in this approach differs
from the standard one as eg.~already the Yang Mills classical
propagator is massive with a mass proportional to the cosmological
constant.  We emphasize that the symmetry preserving scheme employed
here also leads to a vanishing coefficient in this set-up, $I=0$.\step

Based on the general considerations underlying our RG result, we can
acertain that the gauge-fixing and regularisation-dependent
coefficient $I_{\rm 1-loop}\ge 0$.  As long as the implicit or
explicit regularisation respects the symmetry relation
\eq{eq:angularaverage}, the result reads $I=0$.  Note in this context
that regularisations such as Pauli-Villars, or other gauges such as a
general $R_\xi$ gauge, do not respect the symmetry constraint. We also
remark that in the computation of a non-universal coefficient one
cannot tell apart regularisation and gauge-fixing dependences. It is
therefore not surprising that different regularisations and gauges
lead to different coefficients $I\ge 0$.  We conclude that all
existing studies agree in that the gravitational contribution to the
Yang-Mills $\beta$-function support asymptotic freedom.\step

The one-loop approximation is valid in the weak gravity regime.  Close
to the Planck scale and beyond, the dynamics within the gravitational
sector becomes relevant.  We have shown that asymptotic freedom
persists for general anomalous dimension $\eta_N$, which entails its
compatibility with general field-theory-based gravity scenarios. This
includes gravitational shielding as well as asymptotically
safe gravity. \step

In conclusion,  provided that all symmetries and in particular 
\eq{eq:angularaverage} are preserved, the graviton induced  
running of the Yang-Mills coupling vanishes,
\begin{equation}\label{final} 
{\eta_{A}}|_{\rm grav}=0\,. 
\end{equation}
This result stays valid in the presence of a cosmological constant,
and in the presence of a gravitational fixed point.  Hence, it will be
interesting to extend this study to the fully coupled Yang-Mills
gravity system. \step

\section*{Acknowledgments}

This work is supported by Helmholtz Alliance HA216/ EMMI,
and by the Science and Technology Research Council [grant number
ST/G000573/1].

%********|*********|*********|*********|*********|*********|*********|****

\end{document}